\documentclass[aps,twocolumn,pre,longbibliography,unsortedaddress,floatfix]{revtex4-1} 

\usepackage{amsmath,amssymb,amsfonts,amsthm}
\usepackage{mathtools}
\usepackage{bm}
\usepackage[dvipdfmx]{graphicx}
\usepackage{color}
\usepackage{url}
\usepackage{here}

\newcommand{\ddt}{\frac{d}{dt}}
\newcommand{\bc}{\bm{c}}
\newcommand{\bC}{\bm{C}}
\newcommand{\bd}{\bm{d}}
\newcommand{\bD}{\bm{D}}
\newcommand{\base}[1]{\bm{e}_{#1}}
\newcommand{\bF}{\bm{F}}
\newcommand{\bI}{\bm{I}}
\newcommand{\bJ}{\bm{J}}
\newcommand{\bq}{\bm{q}}
\newcommand{\bU}{\bm{U}}
\newcommand{\bX}{\bm{X}}
\newcommand{\bz}{\bm{z}}
\newcommand{\bZ}{\bm{Z}}
\newcommand{\bvsigma}{\boldsymbol{\varsigma}}
\newcommand{\bzeta}{\boldsymbol{\zeta}}
\newcommand{\bcalX}{\boldsymbol{\calX}}
\newcommand{\calX}{\mathcal{X}}
\newcommand{\bbR}{\mathbb{R}}
\newcommand{\Dt}{\Delta t}

\newcommand{\squared}[2]{\left( #1^{2} + #2^{2} \right)}
\newcommand{\ddtheta}{\frac{d}{d\theta}}
\newcommand{\flow}[2]{\Psi_{#1,#2}}
\renewcommand{\mod}[1]{\;\; (\mathrm{mod} \;\; #1)}
\newcommand{\argmax}{\mathop{\rm argmax}\limits}
\newcommand{\argmin}{\mathop{\rm argmin}\limits}
\newcommand{\pd}[2]{\frac{\partial #1}{\partial #2}}

\begin{document}
    \title{Estimating asymptotic phase and amplitude functions of limit-cycle oscillators from time series data}
	
    \author{Norihisa Namura}
    \thanks{Corresponding author. E-mail: namura.n.aa@m.titech.ac.jp}
    \affiliation{Department of Systems and Control Engineering, Tokyo Institute of Technology, Tokyo 152-8552, Japan}

    \author{Shohei Takata}
    \affiliation{Department of Systems and Control Engineering, Tokyo Institute of Technology, Tokyo 152-8552, Japan}

    \author{Katsunori Yamaguchi}
    \affiliation{Department of Systems and Control Engineering, Tokyo Institute of Technology, Tokyo 152-8552, Japan}

    \author{Ryota Kobayashi}
    \affiliation{Graduate School of Frontier Sciences, The University of Tokyo, Chiba 277-8561, Japan,}
    \affiliation{Mathematics and Informatics Center, The University of Tokyo, Tokyo 113-8656, Japan,}
    \affiliation{JST, PRESTO, Saitama 332-0012, Japan}

    \author{Hiroya Nakao}
    \affiliation{Department of Systems and Control Engineering, Tokyo Institute of Technology, Tokyo 152-8552, Japan}

    \date{\today}

    \begin{abstract}
        We propose a method for estimating the asymptotic phase and amplitude functions of limit-cycle oscillators using observed time series data 
        without prior knowledge of their dynamical equations.
        The estimation is performed by polynomial regression and can be solved as a convex optimization problem.
        The validity of the proposed method is numerically illustrated by using two-dimensional limit-cycle oscillators as examples.
        As an application, we demonstrate data-driven fast entrainment with amplitude suppression using the optimal periodic input derived from the estimated phase and amplitude functions.
    \end{abstract}

    \maketitle


    \section{INTRODUCTION}

        There are various nonlinear rhythmic phenomena in the real world, including brain waves~\cite{brain,stankovski2017neural}, animal gaits~\cite{gait1,gait2,kobayashi2016,funato2016evaluation}, heartbeats and respiration~\cite{heart},  passive walking~\cite{hobbelen2007limit, garcia1998simplest}, many of which can be modeled mathematically as limit-cycle oscillators~\cite{Strogatz}.
        The phase reduction method~\cite{Kuramoto1984,Hoppensteadt1997,Winfree2001,Ermentrout2010,Nakao2016,monga2019phase,Kuramoto2019,Ermentrout2019} is useful for analyzing synchronization properties of limit-cycle oscillators subjected to weak perturbations,
        which represents the state of a multidimensional nonlinear oscillator using only the phase variable introduced along its limit cycle and describes the dynamics approximately by a one-dimensional phase equation.
        Recently, the phase reduction method has been extended to the phase-amplitude reduction method~\cite{Wedgwood2013,Wilson2016, mauroy2016global,Shirasaka2017, mauroy2018global, monga2019phase, kotani2020nonlinear,shirasaka2020phase, Nakao2021}, which incorporates the amplitude variable characterizing the distance of the system state from the limit cycle.
        The resulting phase-amplitude equation can be used, for example, in deriving the optimal periodic force for stable entrainment that suppresses amplitude deviations~\cite{Takata2021, shirasaka2020phase, monga2019optimal}.
        
        The phase reduction method is based on the notion of the asymptotic phase~\cite{Winfree2001}, but it is generally not possible to analytically obtain the phase function that gives the asymptotic phase of the system state even if the explicit mathematical model of the oscillator is available.
        Similarly, it is not possible to analytically determine the amplitude functions of the oscillator in general.
        Moreover, if the mathematical model of the target oscillator is unknown, the phase and amplitude functions should be determined from observed data.

        In this study, we propose a simple method for estimating the phase and amplitude functions from the time series data observed from limit-cycle oscillators without relying on their mathematical models.
        Rather than measuring the values of the phase and amplitude by evolving the system state until it converges to the limit cycle, we estimate them by polynomial regression from the differenced time series of the system state started from various initial conditions.
        Our method gives a convex optimization problem, which is computationally inexpensive and can be solved globally.
        We show that the phase and amplitude functions are estimated reasonably well around the limit cycle, including moderately nonlinear regimes, by the proposed method using known models of limit-cycle oscillators as examples.

        For estimating the phase and amplitude functions from observed data,
        Extended Dynamic Mode Decomposition (EDMD)~\cite{schmid2010dynamic,  kutz2016dynamic,Williams2015} and other system-identification methods~\cite{lusch2018deep, wilson2020data, wilson2021data} can also be employed. 
        In particular, EDMD can estimate the Koopman eigenvalues and eigenfunctions of the system from observed data~\cite{Williams2015, klus2020data,li2017extended, takeishi2017learning, alford2021deep, terao2021extended, gulina2021two,kevrekidis2016kernel, kutz2016dynamic, heas2020generalized}.
		The natural frequency and Floquet exponents of the oscillator can then be evaluated from the eigenvalues, and the phase and amplitude functions can be obtained from the associated eigenfunctions~\cite{mauroy2016global,mauroy2018global,Mauroy2020,mauroy2013isostables,Shirasaka2017}.
        In contrast to EDMD, our proposed method estimates the natural frequency and the dominant Floquet exponent separately from the observed data by using a conventional method for Lyapunov exponents and then uses them to estimate the phase and amplitude functions by polynomial regression.
        Our method thus gives an alternative to EDMD and can more robustly estimate the phase and amplitude functions depending on the data.

		As closely related but different problems, methods for estimating the phase response properties of limit-cycle oscillators~\cite{PhysRevLett.94.158101,PhysRevLett.103.024101,doi:10.1143/JPSJ.75.114802,doi:10.7566/JPSJ.86.024009,cestnik2018inferring,funato2016evaluation,NAKAE2010752,ota2011measurement} and for estimating phase coupling functions between interacting oscillators from observed data have been proposed and applied to experimental data~\cite{tokuda2007inferring,kralemann2008phase,brain,duggento2012dynamical,stankovski2017coupling,stankovski2017neural,ota2014direct,onojima2018dynamical,suzuki2018bayesian,arai2021extracting}. 
        In these studies, the phase values of the oscillators are extracted from the time series, e.g., by extracting the spike timing of neural oscillators or by using the Hilbert transform, and then the phase response property of the oscillator or the phase coupling functions between the oscillators are estimated under the assumption that the oscillators are described by the phase model. 
        The phase of the oscillator is typically estimated in the vicinity of the limit cycle and deviation from the limit cycle is not considered; the phase and amplitude functions are generally not introduced explicitly.

        This paper is organized as follows.
        We first outline the notions of asymptotic phase and amplitude in Sec.~\ref{sec2} and then describe the method for estimating the phase and amplitude functions by polynomial regression in Sec.~\ref{sec3}.
        The validity of the proposed method is illustrated by numerical simulations using two types of oscillators and how to evaluate the estimated results is described in Sec.~\ref{sec4}.
        Section~\ref{sec5} demonstrates data-driven fast entrainment of the oscillator with amplitude suppression using the estimated phase and amplitude functions, and Sec.~\ref{sec6} gives concluding remarks.


    \section{PHASE AND AMPLITUDE of limit-cycle oscillators}
    \label{sec2}
        

        \subsection{Asymptotic phase and amplitude functions}

            We first introduce the asymptotic phase and amplitude functions~\cite{Wedgwood2013,Wilson2016,Shirasaka2017,Nakao2021}.
            We consider a limit-cycle oscillator described by
            \begin{align}
                \ddt \bX(t) = \bF(\bX(t)),
                \label{eq1}
            \end{align}
            where $\bX(t) \in \bbR^{N}$ is the system state at time $t$.
            We assume that the system has an exponentially stable limit cycle trajectory $\bX_{0}(t)$ with a natural period $T$ and frequency $\omega = 2\pi/T$, which is a $T$-periodic function of $t$ satisfying $\bX_{0}(t + T) = \bX_{0}(t)$.
            
            First, we assign a phase $\theta \in [0, 2\pi)$ for each state on the limit cycle, where $0$ and $2\pi$ are considered identical,
            by choosing a state $\bX_{0}(0)$ on the limit cycle as the phase origin, $\theta=0$, and defining the phase of the state $\bX_{0}(t)$ at $t>0$ as $\theta = \omega t \mod{2\pi}$. 
            We denote by $\bcalX_{0}(\theta)$ the state on the limit cycle with phase $\theta$.
            Next, we extend the definition of the phase to the basin of the limit cycle.
            The phase of a system state $\bX_{A}$ in the basin is defined as $\theta$
            if $\lim_{t \to \infty} \left\| \flow{t}{t_{0}}\bX_{A} - \flow{t}{t_{0}}\bcalX_{0}(\theta) \right\| = 0$ holds, 
            where $\flow{t}{t_{0}}$ is the flow of Eq.~(\ref{eq1}) and $\| \cdot \|$ is the $L_2$ norm. 
            That is, if the state $\flow{t}{t_{0}} \bX_{A}$ started from $\bX_{A}$ at $t_0$ converges to the state $\flow{t}{t_{0}} \bcalX_{0}(\theta)$ started from $\bcalX_{0}(\theta)$ at $t_0$ as $t \to \infty$, we consider that the phase state $\bX_{A}$ has the same phase $\theta$ as $\bcalX_{0}(\theta)$.
            This defines a phase function $\Theta$ for all states $\bX$ in the basin, which assigns a phase value $\theta(t) = \Theta(\bX(t))$ to the state $\bX(t)$ and satisfies
            \begin{align}
                \ddt \theta(t) = \ddt \Theta(\bX(t)) = \omega.
                \label{eq:omega}
            \end{align}
            The phase defined in this way is called the asymptotic phase,
            and the level sets of the phase function is called ``isochrons''~\cite{Kuramoto1984,Hoppensteadt1997,Winfree2001,Ermentrout2010,Nakao2016}.
            The asymptotic phase can also be interpreted as the argument of the eigenfunction of the Koopman generator $\bF(\bX)\cdot \nabla$ of Eq.~(\ref{eq1}) associated with the eigenvalue $i\omega$~\cite{mauroy2016global,mauroy2018global,Mauroy2020,mauroy2013isostables,Shirasaka2017}.

            Next, we introduce the (dominant) amplitude function.
            In a similar way to the phase $\theta(t)$, we assign a scalar amplitude $r(t) = R(\bX(t))$ for all states $\bX$ the basin, where the amplitude function $R$ satisfies 
            \begin{align}
                \label{eq:R}
                \ddt r(t) = \ddt R(\bX(t)) = \lambda R(\bX(t)) = \lambda r(t).
            \end{align}
            Here, the coefficient $\lambda < 0$ is the Floquet exponent of the limit cycle with the largest non-zero real part, which is assumed to be real and simple.
			As the state $\bX(t)$ converges to the limit cycle, the amplitude $r(t)$ decays exponentially and satisfies $R(\bcalX_{0}(\theta)) = 0$ when the state is on the limit cycle.
            The amplitude function can also be considered an eigenfunction of the Koopman generator $\bF(\bX)\cdot \nabla$ of Eq.~(\ref{eq1}) associated with the eigenvalue $\lambda$.
			The level sets of the amplitude function is called ``isostables'' in the framework of the Koopman operator analysis of limit-cycling systems~\cite{mauroy2016global,mauroy2018global,Mauroy2020,mauroy2013isostables,Shirasaka2017}.

			By using the phase and amplitude functions, we can reduce the dimensionality of the oscillator and globally linearize the dynamics in the basin of the limit cycle, which yields a simple description of the oscillator useful for the analysis. 
            Moreover, near the limit cycle, we can write down phase-amplitude equations describing weakly perturbed oscillatory dynamics. 
            The simplicity of the phase equation has played a prominent role in the theoretical analysis of synchronization phenomena in various types of limit-cycling systems~\cite{Kuramoto1984,Hoppensteadt1997,Winfree2001,Ermentrout2010}.
          

        \subsection{Response and sensitivity functions}
        \label{sec2B}

            We next introduce the response and sensitivity functions of the phase and amplitude, which are used for the validation of our estimation method. 

            The phase response function (PRF, also known as the phase response or resetting curve, PRC~\cite{Kuramoto1984,Winfree2001,Brown2004,Ermentrout2010}) 
            gives the phase difference of the oscillator caused by an impulse perturbation applied to the oscillator at phase $\theta$ and is defined as
            \begin{align}
                g(\theta;\bvsigma) &= \Theta(\bcalX_{0}(\theta) + \bvsigma) - \Theta(\bcalX_{0}(\theta)) = \Theta(\bcalX_{0}(\theta) + \bvsigma) - \theta,
            \end{align}
            where $\bvsigma \in \bbR^{N}$ represents the direction and intensity of the impulse~\cite{Winfree2001,Nakao2005,Ermentrout2010,Nakao2016,arai2008phase}.
            When $|\bvsigma|$ is sufficiently small, we can approximate $g$ by using the Taylor expansion of $\Theta$ near $\bvsigma = \bm{0}$, 
            $\Theta(\bcalX_{0}(\theta)~+~\bvsigma) = \Theta(\bcalX_{0}(\theta)) + \nabla \Theta(\bX)|_{\bX = \bcalX_{0}(\theta)}~\cdot~\bvsigma +~O(|\bvsigma|^{2})$, as {$g(\theta ; \bvsigma) \simeq \nabla \Theta(\bX)|_{\bX = \bcalX_{0}(\theta)} \cdot \bvsigma$}.
            We denote the gradient of $\Theta(\bX)$ evaluated at the state $\bX = \bcalX_{0}(\theta)$ by
            \begin{align}
                \bZ(\theta) = \nabla \Theta(\bX)|_{\bX = \bcalX_{0}(\theta)}
            \end{align}
            and call it the phase sensitivity function (PSF, also known as the infinitesimal phase resetting curve, iPRC). 
            This $\bZ(\theta)$ characterizes the linear response of the oscillator phase to a weak perturbation given at phase $\theta$.
            
            Similarly, we introduce the amplitude response function (ARF) 
            characterizing the amplitude difference of the oscillator caused by an impulse $\bvsigma$ given at phase $\theta$ as
            \begin{align}
                h(\theta;\bvsigma) &= R(\bcalX_{0}(\theta) + \bvsigma) - R(\bcalX_{0}(\theta))
                = R(\bcalX_{0}(\theta) + \bvsigma).
            \end{align}
            When $|\bvsigma|$ is sufficiently small, we can approximate $h$ by using the Taylor expansion of $R(\bcalX_{0}(\theta) + \bvsigma)$ near $\bvsigma = \bm{0}$,
            {$R(\bcalX_{0}(\theta) + \bvsigma) = R(\bcalX_{0}(\theta)) + \nabla R(\bX)|_{\bX = \bcalX_{0}(\theta)} \cdot \bvsigma + O(|\bvsigma|^{2})$}, as $h(\theta;\bvsigma) \simeq {\nabla R(\bX)|_{\bX = \bcalX_{0}(\theta)}} \cdot \bvsigma$.
            Here, we denote the gradient of the amplitude function $R(\bX)$ at the state $\bX = \bcalX_{0}(\theta)$ by
            \begin{align}
                \bI(\theta) = \nabla R(\bX)|_{\bX = \bcalX_{0}(\theta)}
            \end{align}
            and call it the amplitude sensitivity function (ASF, also known as isostable response function in the Koopman operator analysis of limit-cycling systems).
            This $\bI(\theta)$ characterizes the linear response of the oscillator amplitude to a weak perturbation given at $\theta$.

            If the mathematical model of the oscillator is known, the PSF $\bZ(\theta)$ and ASF $\bI(\theta)$ can be determined by numerically calculating the $2\pi$-periodic solutions of the following adjoint equations~\cite{Brown2004,Ermentrout2010,Kuramoto2019,Shirasaka2017,Takata2021}:
            \begin{align}
				\label{eq:adjZ}
                \omega \ddtheta \bZ(\theta) &= - \bJ(\theta)^{\top} \bZ(\theta), \\
				\label{eq:adjI}
                \omega \ddtheta \bI(\theta) &= - \left( \bJ(\theta)^{\top} - \lambda \right) \bI(\theta),
            \end{align}
            where $\bJ(\theta)$ denotes the Jacobian matrix of the vector field $\bF$ at $\bX = \bcalX_{0}(\theta)$.
            The PSF $\bZ(\theta)$ should satisfy $\bZ(\theta) \cdot \bF \left(\bcalX_{0}(\theta) \right) = \omega$ as a normalization condition. 
            The normalization condition for the ASF can be chosen arbitrary and will be specified later.


        \subsection{Direct numerical calculation of phase and amplitude functions}
        
            It is generally difficult to obtain the phase and amplitude functions analytically from a mathematical model, but they can be obtained by direct numerical calculation of the mathematical model as follows.

            For the phase function~\cite{Nakao2016}, we first choose a point on the limit cycle as the origin of the phase, then assign the phase values to the states on the limit cycle.
            For the states in the basin of the limit cycle, we let them evolve for an integer multiple of the period $T$ until they converge to the limit cycle and then determine their phase values.

            Similarly, for the amplitude function, we let the state $\bX$ in the basin evolve until it becomes sufficiently close to (but not completely on) the limit cycle, and find the point $\bcalX(\theta)$ on the limit cycle with the same phase $\theta$ as $\bX$. 
            The amplitude of the point sufficiently close to the limit cycle can be obtained by taking the inner product of the vector $\Delta \bX = \bX - \bcalX(\theta)$ between the two points and the amplitude sensitivity function $\bI(\theta)$ as $R(\bX) \simeq \bI(\theta) \cdot \Delta \bX$, which follows from the Taylor expansion of the amplitude function.

            Note that these methods are exhaustive and computationally required. We will use the phase and amplitude functions directly calculated from the mathematical model by the above method as `true' functions to characterize the accuracy of the estimated functions from the observed data.


    \section{Proposed method of estimation}
    \label{sec3}
    

        \subsection{Estimation of the frequency and Floquet exponent}
    
            In our method, the natural frequency $\omega$ and the largest non-zero Floquet exponent $\lambda$ of the oscillator should be measured before the estimation of the phase and amplitude functions.
            The natural frequency $\omega$ can be estimated from the average period $T$ of the system state to perform rotations.
            To estimate the sum of the Floquet exponent from time series data, we use the method of Wolf~{\it et~al.~}\cite{lyapunov} for the Lyapunov exponents of dynamical systems.
            For two-dimensional limit-cycle oscillators, 
            the Floquet exponents are real and exactly the same as the Lyapunov exponents.
            As one Lyapunov exponent $\lambda_1$ corresponding to the phase direction is $0$, it is enough to estimate the sum of the Lyapunov exponents, $\lambda_1+\lambda_2$, which can be performed by measuring the  evolution of a small phase-plane area for a long time.
			The same method can also be used for the two largest Lyapunov exponents of higher-dimensional oscillators when $\lambda$ is real. 

            We assume that a discretized time sequence of the system state (with observation noise) from an initial condition not on the limit cycle is obtained. 
            By taking the logarithmic ratio of the area $S(t_{k})$ of the triangle formed by three nearby states in the time series at time $t_{k}$ to the area $S(t_{k+1})$ at $t_{k+1}$ and summing over a long period of time, we can obtain $\lambda$ as the sum of the Lyapunov exponents:
            \begin{align}
                \label{eq:lyapunov}
                \lambda = \lambda_{1} + \lambda_{2} = \frac{1}{t_{L} - t_{0}} \sum_{k=0}^{L-1} \log \frac{S(t_{k+1})}{S(t_{k})},
            \end{align}
            where $L$ is the length of the data and the area $S(t_{k})$ is reset to $0.1$ at each $t_{k}$ so that it does not vanish due to numerical errors.


        \subsection{Estimation of the phase function}
        \label{sec3B}
        
            We first propose a method for estimating the phase function by polynomial regression from observed data of an unknown limit-cycle oscillator.
            We do not measure the absolute phase value of each state, which is difficult with the time series; rather, we use the phase differences between two consecutive states in the time series and use Eq.~\eqref{eq:omega} for the estimation.

            We assume that we can obtain discretized time sequences of $\bX(t) \in \bbR^{N \times 1}$ from various initial conditions (with observation noise). 
            Each sequence is sampled at equal intervals of $\Dt$ with {$M$} sampling points, where $\Dt$ is sufficiently smaller than the natural period $T$ of the oscillator.
            We construct a standardized row vector of polynomials up to order $p \geq 1$ from $\bX$,
            \begin{align}
                \label{eq:U}
                &\bU(\bX,p) = \nonumber \\ & \quad
                \begin{bmatrix}
                    1  & \bar{x}_1  &  \cdots  &  \bar{x}_N  & \bar{x}^2_1 &  \overline{ x_1 x_2} &   \cdots   &  \bar{x}_N^2  &   \cdots   &  \bar{x}_N^p
                \end{bmatrix}
                \in \bbR^{1 \times P},
            \end{align}
			where $x_k$ is the $k$-th component of $\bX$ and all cross terms like $\overline{x_1^2 x_2^3}$ up to order $p$ are included.
            The overline indicates standardization, i.e., $\bar{z}=: (z- \mu)/\sigma$ is the standard score of $z$,
            where $\mu$ and $\sigma$ are the mean and the standard deviation of $z$, respectively. 
            We consider approximating the sine and cosine of the phase function $\Theta(\bX)$ by polynomials of order $p$ as 
            \begin{align}
                \label{eq:poly_approx}
                \begin{aligned}
                    \cos(\Theta(\bX)) \simeq \bU(\bX,p)\bz_{1}, \quad 
                    \sin(\Theta(\bX)) \simeq \bU(\bX,p)\bz_{2},
                \end{aligned}
            \end{align}
            where $\bz_{1} \in \bbR^{P \times 1}$ and $\bz_{2} \in \bbR^{P \times 1}$ are the coefficient vectors of the polynomials.
            Hereafter, $\bU(\bX,p)$ is denoted by $\bU(\bX)$.
            Our aim is to find the best polynomial approximations of $\sin \Theta$ and $\cos \Theta$ that are consistent with the definition of $\Theta$, Eq.~(\ref{eq:omega}), from the time series of $\bX(t)$.
                    
            Taking the time derivatives of $\sin \Theta(\bX)$ and $\cos \Theta(\bX)$, we obtain
            \begin{align}
                \begin{aligned}
                    \ddt\cos(\Theta(\bX(t))) &= - \omega \sin(\Theta(\bX(t))),\\
                    \ddt\sin(\Theta(\bX(t))) &= \omega \cos(\Theta(\bX(t))),
                \end{aligned}
            \end{align}
            and using Eq.~\eqref{eq:poly_approx}, we obtain 
            \begin{align}
				\label{eq:14}
                \begin{aligned}
                    \ddt \bU(\bX(t)) \bz_{1} &= \left( \left( \left. \pd{\bU(\bX)}{\bX} \right|_{\bX=\bX(t)} \right)^{\top} \frac{d\bX(t)}{dt} \right) \cdot \bz_{1} \\
                    &\simeq - \omega \bU(\bX(t)) \bz_{2}, \\
                    \ddt \bU(\bX(t)) \bz_{2} &= \left( \left( \left. \pd{\bU(\bX)}{\bX} \right|_{\bX=\bX(t)} \right)^{\top} \frac{d\bX(t)}{dt} \right) \cdot \bz_{2} \\
                    &\simeq \omega \bU(\bX(t)) \bz_{1},
                \end{aligned}
            \end{align}
            where $\partial {\bU(\bX)} / \partial {\bX} \in \bbR^{N \times P}$ is a matrix whose $(j, k)$-component is given by {$\partial U_k / \partial X_j$} ($j=1, \dots, N, k=1, \dots, P$) and ``$\cdot$'' indicates the dot product of two vectors. 

            We discretize the time as $t = i \Dt$ with a small time step $\Dt$ and represent the time series $\bX(t)$ as $\bX_i = \bX(i \Dt)$. We also denote numerical derivative of $\bX(t)$ at $t=i\Delta t$ as $\dot\bX_i$, which is calculated by a linear regression of several consecutive data points in the time series since observation noise is  overlapped on the data~\cite{polyinter}.     
            From Eq.~(\ref{eq:14}), the following equations should hold approximately for each $i=1, \dots, M$: 
            \begin{align}
                \begin{aligned}
                     { \left( \left. \pd{\bU(\bX)}{\bX} \right|_{\bX = \bX_{i}} \right)^{\top}}  \dot\bX_i \bz_{1} + \omega \bU(\bX_{i}) \bz_{2} &= 0, \\
                    { \left( \left. \pd{\bU(\bX)}{\bX} \right|_{\bX = \bX_{i}} \right)^{\top}} \dot\bX_i \bz_{2} - \omega \bU(\bX_{i}) \bz_{1} &= 0.
                \end{aligned}
            \end{align}
            Introducing $\bz = \left[ \bz_{1}^{\top},\bz_{2}^{\top} \right]^{\top} \in \bbR^{2P \times 1}$, these equations can be expressed as 
            \begin{align}
                \begin{aligned}
                    \begin{bmatrix}
                        (\bU'_{i})^{\top} \dot\bX_i
                        & \omega \bU_i \\ 
                        - \omega \bU_i
                        &
                        (\bU'_i)^{\top} \dot\bX_i
                    \end{bmatrix}
                    \bz \coloneqq{}& \bc_{i}\bz
                    ={}
                    \begin{bmatrix}
                        0 \\ 0
                    \end{bmatrix}
                    ,
                \end{aligned}
            \end{align}
            where $\bU_i = \bU(\bX_i)$ and $\bU'(\bX_{i}) = \partial U(\bX)/\partial \bX|_{\bX = \bX_i}$.
            We try to find the best coefficient vector $\bz$ that satisfies the above equation as much as possible for the given time series.
            Defining a matrix $\bC$ by $\bC = \left[ \bc_{1}^{\top}, \bc_{2}^{\top}, \cdots, \bc_{M}^{\top} \right]^{\top}\in \bbR^{2M \times 2P}$, this gives a minimization problem of the following objective function:
            \begin{align}
                E_{\Theta} = \| \bC \bz \|^{2}
            \end{align}
            where $\| \bC\bz \|^{2} = \sum_{k=1}^{2M} \{( \bC\bz )_k\}^2$.
            
            To fix the origin of the estimated phase function, we also require that the absolute phase of a given state $\tilde{\bX}$ is zero, i.e.,
            \begin{align}
                \Theta\left(\tilde{\bX}\right) = 0.
            \end{align}
            Thus, the polynomial approximations of $\cos \Theta\left( \tilde{\bX} \right)$ and $\sin \Theta\left( \tilde{\bX} \right)$ should satisfy
            \begin{align}
                \begin{aligned}
                    \bU\left( \tilde{\bX} \right)\bz_{1} = 1, \quad 
                    \bU\left( \tilde{\bX} \right)\bz_{2} = 0,
                \end{aligned}
            \end{align}
            and using $\bz = \left[ \bz_{1}^{\top},\bz_{2}^{\top} \right]^{\top}$, 
            \begin{align}
                \begin{bmatrix}
                    \bU\left( \tilde{\bX} \right) & \bm{0} \\
                    \bm{0} & \bU\left( \tilde{\bX} \right)
                \end{bmatrix}
                \bz = 
                \begin{bmatrix}
                    1 \\ 0
                \end{bmatrix}
                .
            \end{align}

            Considering the above objective function and the constraints, the proposed estimation method can be formulated as an optimization problem for the coefficient $\bz$ as follows:
            \begin{align}
                \begin{aligned}
                    &&\hat{\bz} ={} &\argmin_{\bz} \| \bC\bz \|^{2} \\
                    &&&\mathrm{s.t.} 
                    \begin{bmatrix}
                        \bU\left( \tilde{\bX} \right) & \bm{0} \\
                        \bm{0} & \bU\left( \tilde{\bX} \right)
                    \end{bmatrix}
                    \bz = 
                    \begin{bmatrix}
                        1 \\ 0
                    \end{bmatrix}
                    .
                \end{aligned}
            \end{align}
            This optimization problem is convex with respect to the coefficient vector $\bz$. 


        \subsection{Estimation of the amplitude function}
        \label{sec3C}

            We next develop a method for estimating the amplitude function by polynomial regression using the same observed data as in the previous subsection.
            For simplicity, we consider the amplitude function {of} a two-dimensional oscillator; the method can also be applied straightforwardly to higher-dimensional oscillators whose Floquet exponent with the largest non-zero part is real.
            
            Since the Floquet exponents of a two-dimensional oscillator are real, i.e., a zero exponent associated with the phase (tangential) direction and another negative exponent associated with the amplitude direction, the amplitude function is also real.
            Therefore, we can approximate the amplitude function $R(\bX)$ by a polynomial as 
            \begin{align}
                \label{eq:amp_poly}
                R(\bX) \simeq \bU(\bX) \bzeta,
            \end{align}
            where $\bzeta \in \bbR^{P \times 1}$ is the coefficient vector.
            Plugging Eq.~\eqref{eq:amp_poly} into Eq.~(\ref{eq:R}), we obtain
            \begin{align}
                \left( \left( \left. \pd{\bU(\bX)}{\bX} \right|_{\bX=\bX(t)} \right)^{\top} \frac{d\bX(t)}{dt} \right) \cdot \bzeta = \lambda \bU(\bX) \bzeta.
            \end{align}
            Discretizing the time and plugging the time series data, we obtain
            \begin{align}
                \begin{bmatrix}
                    \left( \left. \pd{\bU(\bX)}{\bX} \right|_{\bX = \bX_{i}} \right)^{\top} \dot\bX_i - \lambda \bU(\bX_{i}) 
                \end{bmatrix}
                \bzeta \coloneqq \bd_{i} \bzeta = 0
            \end{align}
            for $i=1, ..., M$, where $\bd_i \in \bbR^{1 \times P}$. 
            As in the case for the phase function, the numerical differentiation of $\bX(t)$ is evaluated by the polynomial interpolation.
            Defining a matrix $\bD$ by $\bD = \left[ \bd_{1}^{\top}, \bd_{2}^{\top}, \cdots, \bd_{M}^{\top} \right]^{\top} \in \bbR^{M \times P}$, we obtain the following objective function to minimize:
            \begin{align}
                E_{R} =  \| \bD\bzeta \|^{2}.
            \end{align}
    
            On the limit cycle, the value of the amplitude function should take $0$, i.e., $R(\bcalX_{0}(\theta)) = 0$.
            This condition is given as the constraint for the polynomial approximation of $R$ as $\bU(\bX_{0,i}) \bzeta = 0$, where $\bX_{0,i}$ $(i=1, ..., M_0)$ denotes $M_0$ oscillator states on the limit cycle, which can be chosen at equal time intervals. 
            Defining a matrix $\bU_{0}$ by $\bU_{0} = \left[ \bU(\bX_{0,1})^{\top}, \bU(\bX_{0,2})^{\top}, \cdots, \bU(\bX_{0,M_{0}})^{\top} \right]^{\top}$, we obtain
            \begin{align}
                \bU_{0} \bzeta = \bm{0}.
            \end{align}
    
            As stated previously, the scale of the amplitude can be arbitrarily chosen. 
            To fix the scale of the amplitude function (this also fixes the scale of the amplitude sensitivity function $\bI(\theta)$ introduced in Sec.~\ref{sec2B}, we specify the absolute amplitude value of a given state $\tilde{\bX}$ as $R \left( \tilde{\bX} \right) = r_{0}$ ($r_0>0$)
            as in the case of the phase function.
            This point $\tilde{\bX}$ should not be on the limit cycle and should be chosen near the {fixed point} of the system in order to capture the shape of the amplitude function appropriately. 
            This constraint can be expressed as
            \begin{align}
                \bU \left( \tilde{\bX} \right) \bzeta \coloneqq \tilde{\bU} \bzeta = r_{0}.
            \end{align}
            
            Moreover, we require that the norm of the coefficient vector $\bzeta$ does not become too large so that the estimated amplitude function does not diverge outside of the limit cycle due to the above two constraints near the {fixed point} and on the limit cycle.
            We thus include an additional objective function proportional to the norm $\| \bzeta \|^2$.
    
            Summarizing, the proposed estimation method can be formulated as an optimization problem for the coefficient vector $\bzeta$ as follows:
            \begin{align}
                \label{eq:amp_formula}
                \begin{aligned}
                    &&\hat{\bzeta} ={} &\argmin_{\bzeta} \left\| \bD \bzeta \right\|^{2} + \gamma \| \bzeta \|^{2}\\
                    &&&\mathrm{s.t.} 
                    \begin{bmatrix}
                        \tilde{\bU} \\ \bU_{0}
                    \end{bmatrix}
                    \bzeta = 
                    \begin{bmatrix}
                        r_{0} \\ \bm{0}
                    \end{bmatrix},
                \end{aligned}
            \end{align}
            where $\gamma$ is a parameter for the penalty for the solution norm.
            This optimization problem is convex with respect to the coefficient vector $\bzeta$.

			The second term of the objective function in Eq.~\eqref{eq:amp_formula} takes the same form as the regularization term in the ridge regression. 
            The hyper parameter $\gamma$ was determined by using the L-curve~\cite{Lcurve}. 
            The L-curve is defined by a log-log plot of the error $\|D\bzeta\|^{2}$ and the solution norm $\|\bzeta\|^{2}$. 
            We empirically choose the value of $\gamma$ where the absolute value of the slope falls below $6$.


    \section{Verification of the proposed method}

    \label{sec4}


        \subsection{Data used for estimation}
        
            In this section, we verify the validity of the proposed method by numerical simulations using two-dimensional limit-cycle oscillators, the Stuart-Landau oscillator~\cite{Kuramoto1984, Nakao2016} and the van der Pol oscillator~\cite{van1927vii, van1927frequency, van1926lxxxviii}. 
            The time series data are produced by numerical integration of the dynamical system 
            from $n$ initial states that are taken uniformly at random in a certain region, where each initial state is evolved until it converges to the limit cycle and $M$ data points are sampled from the trajectory.
            Here, we took $n$ initial points to increase the number of data points outside the limit cycle, while we assumed a single time series in our explanation of the methods in Sec.~\ref{sec3}.
            These $n$ time series are concatenated and treated as one time series vector, but the regression and differentiation are performed only within the individual time series.


        \subsection{Evaluation of estimated results}
        
                To examine the performance of the proposed method, we consider limit-cycle oscillators whose mathematical models are known and compare the estimated results with the true results obtained by direct numerical calculations of the mathematical models.
                We evaluate the estimation accuracy by comparing the PSFs, PRFs, ASFs, and ARFs obtained from the estimated and the true phase and amplitude functions.
                In what follows, we use normalized PRFs (nPRFs) and normalized ARFs (nARFs) defined as the PRFs and ARFs divided by the impulse intensity, respectively.

                The $j$-th component of the PSF $\bZ(\theta)$ and ASF $\bI(\theta)$ are obtained by numerical derivative of the estimated phase and amplitude functions as
                \begin{align}
					\label{eq:estZ}
                    \hat{Z}_{j}(\theta_{i}) = \frac{\hat{\Theta}(\bX_{0,i} + \epsilon \base{j}) - \hat{\Theta}(\bX_{0,i} - \epsilon \base{j})}{2\epsilon}
                \end{align}
                and
                \begin{align}
					\label{eq:estI}
                    \hat{I}_{j}(\theta_{i}) = \frac{\hat{R}(\bX_{0,i} + \epsilon \base{j}) - \hat{R}(\bX_{0,i} - \epsilon \base{j})}{2\epsilon},
                \end{align}
                for sufficiently small $\epsilon$, where $\base{j}$ is the unit vector in the $j$-th direction and $\bX_{0,i}$ is the system state on the limit cycle with phase $\theta_i$. We use $\epsilon = 10^{-5}$ in what follows.

                The nPRF 
                ${G}_{\varsigma,j}(\theta) \coloneqq g(\theta, \varsigma \base{j}) / \varsigma$
				and the nARF 
                $H_{\varsigma,j}(\theta) \coloneqq h(\theta, \varsigma \base{j}) / \varsigma$
                of the oscillator with respect to the impulse of finite intensity $\varsigma$ given to the $j$-th component of the system state can be calculated from the estimated phase and amplitude functions as
                \begin{align}
					\label{eq:estG}
                    \hat{G}_{\varsigma,j}(\theta_{i}) 
                    \coloneqq 
                    \frac{\hat{\Theta}(\bX_{0,i} + \varsigma \base{j}) - \hat{\Theta}(\bX_{0,i})}{\varsigma},
                \end{align}
                and
                \begin{align}
					\label{eq:estH}
                    \hat{H}_{\varsigma,j}(\theta_i) \coloneqq \frac{\hat{R}(\bX_{0,i} + \varsigma \base{j}) - \hat{R}(\bX_{0,i})}{\varsigma},
                \end{align}
                respectively.

                We evaluate the accuracy of the PSFs, ASFs, nPRFs, and nARFs by using the coefficient of determination as follows:
                \begin{align}
                    R_{Y}^{2} = 1 - \frac{ \sum_{i} \left( Y_{j}(\theta_{i}) - \hat{Y}_{j}(\theta_{i}) \right)^{2} }
                    { \sum_{i} \left( Y_{j}(\theta_{i}) - \overline{Y}_{j} \right)^{2} },
                \end{align}
                where $Y_j$ is either of $Z_j$, $I_j$, $G_{\varsigma,j}$, or $H_{\varsigma,j}$, and $\overline{Y}_{j}$ represents the mean of $\{ Y_{j}(\theta_i) \}$ over $i$.
                The closer the coefficient of determination is to $1$, the higher the accuracy of the estimation.


            \begin{figure}[tb]
                \centering
                \includegraphics[width=\hsize]{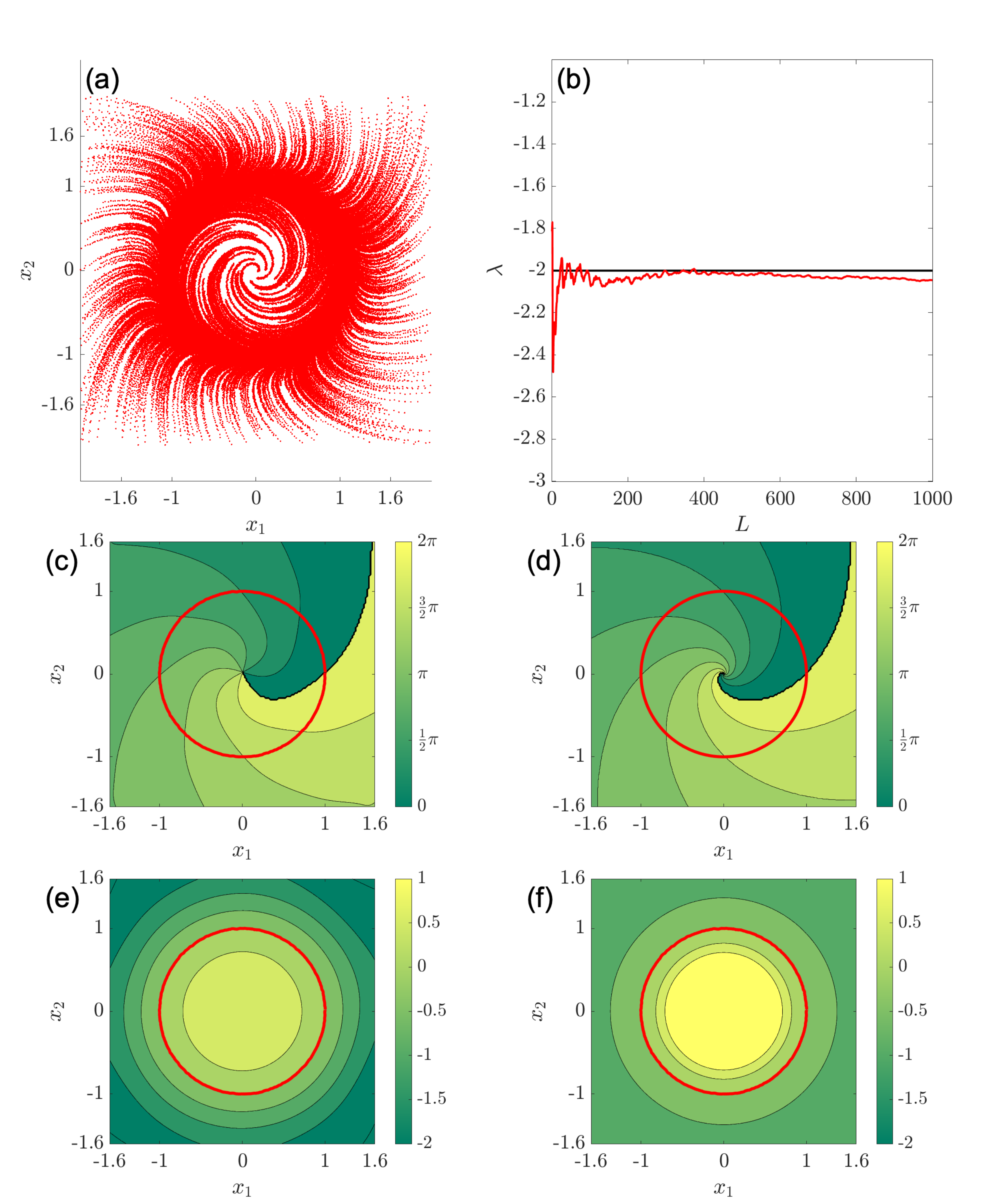}
                \caption{
                    Phase and amplitude functions of the SL oscillator used for the estimation.
                    (a)~Time series data used for the estimation.
                    (b)~Estimated non-zero Floquet exponent.
                    (c)~Estimated phase function.
                    (d)~True phase function.
                    (e)~Estimated amplitude function.
                    (f)~True amplitude function.
                    In (c)-(f), the red bold circle shows the limit cycle.
                }
                \label{fig1}
            \end{figure}

            \begin{figure}[tb]
                \centering
                \includegraphics[width=\hsize]{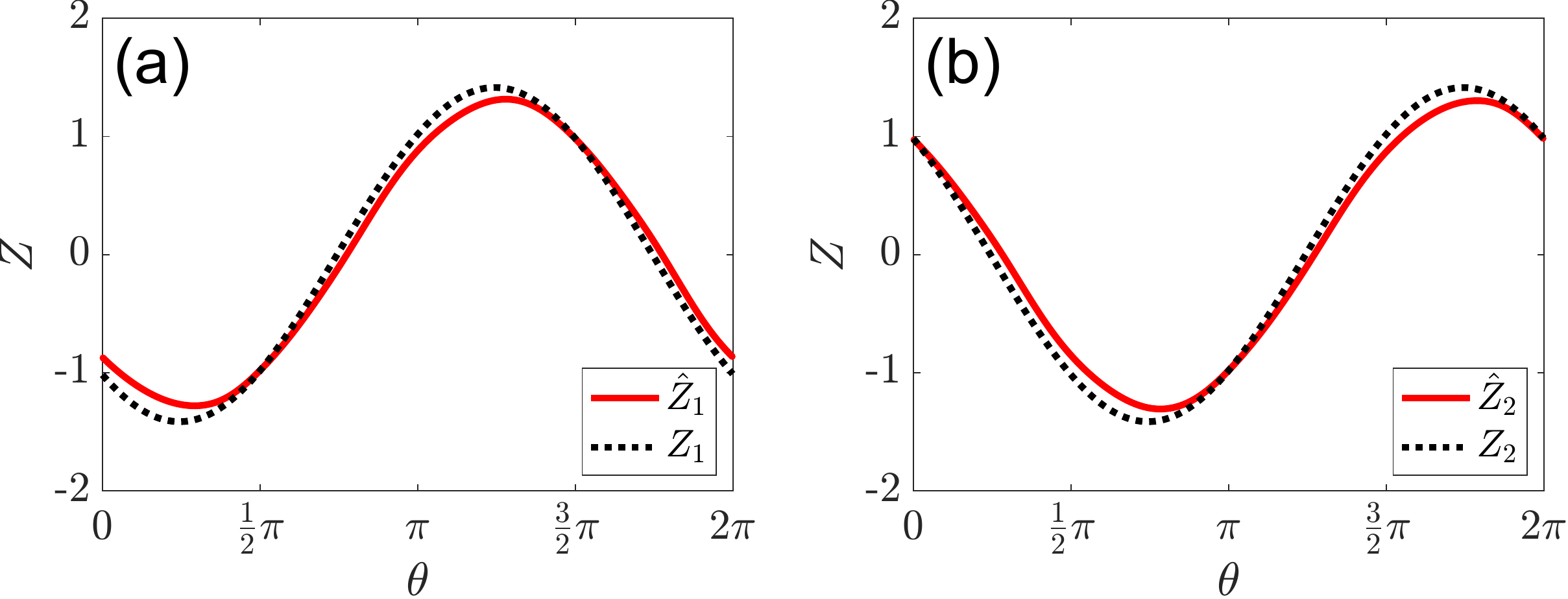}
                \caption{
                    Phase sensitivity functions (PSFs) of the SL oscillator.
                    Each graph shows the estimated PSF (red solid curve) and the true PSF (black dotted curve).
                    (a): $x_1$ component.
                    (b): $x_2$ component.
                    Estimated PSFs are obtained by numerical differentiation with $\epsilon=10^{-5}$. True PSFs are calculated by solving the adjoint equation.
                }
                \label{fig2}
            \end{figure}

            \begin{figure}[tb]
                \centering
                \includegraphics[width=\hsize]{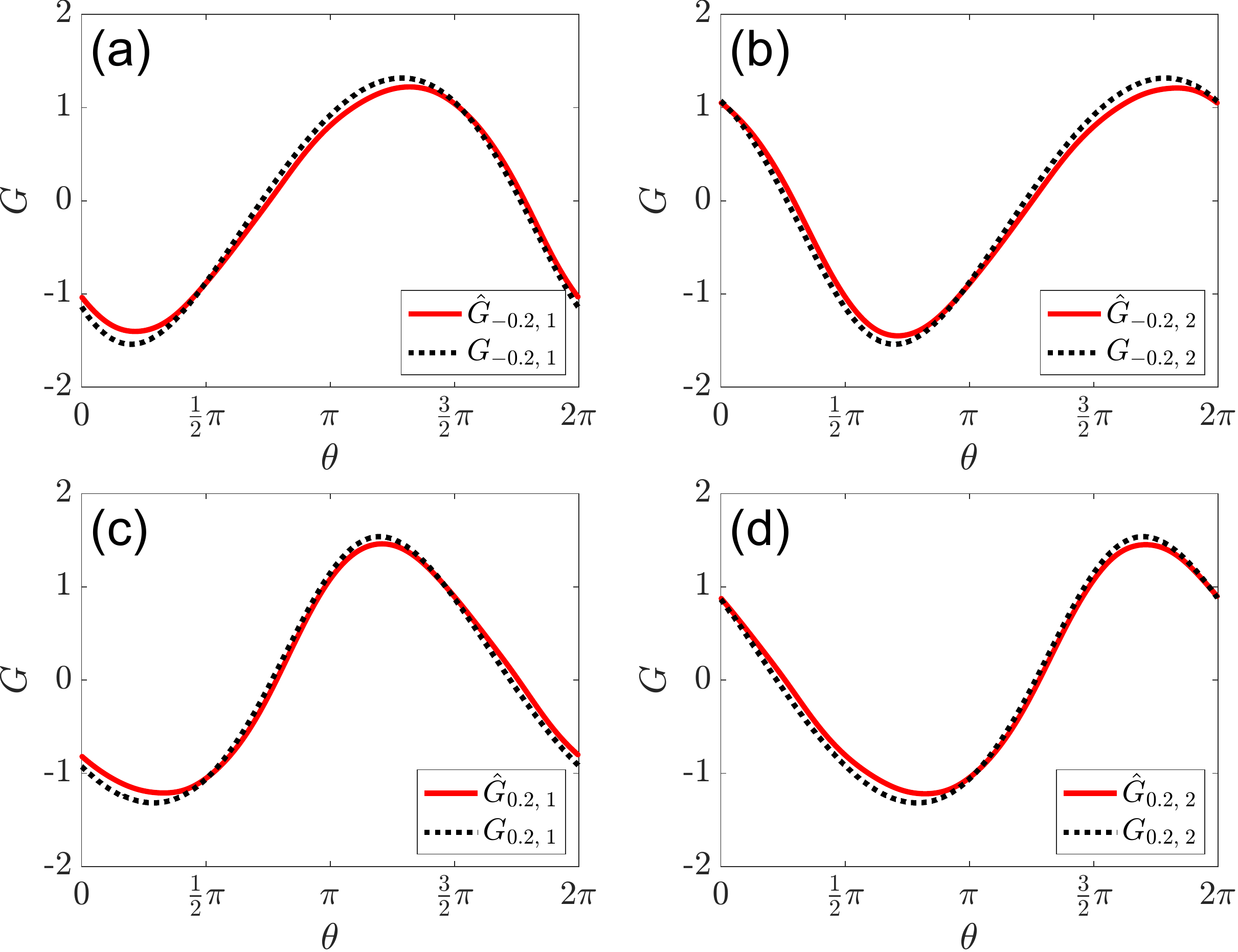}
                \caption{
                    Normalized phase response functions (nPRFs) of the SL oscillator.
                    Each graph shows the estimated nPRF (red solid curve) and the true nPRF (black dotted curve).
                    (a,c):~Impulses applied in the $x_1$ direction.
                    (a)~$\varsigma = -0.2$,
                    (c)~$\varsigma = +0.2$.
                    (b,d):~Impulses applied in the $x_2$ direction.
                    (b)~$\varsigma = -0.2$,
                    (d)~$\varsigma = +0.2$.
                }
                \label{fig3}
            \end{figure}


        \subsection{Stuart-Landau oscillator}

            First, we verify the validity of the proposed method using the Stuart-Landau (SL) oscillator, for which the analytical solutions of the phase and amplitude functions are obtained~\cite{Kato2021asymptotic}.
			The SL oscillator is described as follows:
            \begin{align}
                \ddt
                \begin{bmatrix}
                    x_{1} \\ x_{2}
                \end{bmatrix}
                = 
                \begin{bmatrix}
                    x_{1} - \alpha x_{2} - (x_{1} - \beta x_{2})\squared{x_{1}}{x_{2}} \\
                    \alpha x_{1} + x_{2} - (\beta x_{1} + x_{2})\squared{x_{1}}{x_{2}}
                \end{bmatrix}
                ,
                \label{eq:sl}
            \end{align}
            where $x_1, x_2$ are the variables and $\alpha = 2,\; \beta = 1$ are parameters.
            We set the sampling interval as $\Dt = 0.005$, the number of initial points as $n = 1200$, and the number of data points as $M = 500$.
            We estimated the phase function for the SL oscillator under the observation noise,
            which is given by Gaussian noise with mean $0$ and standard deviation $5 \times 10^{-3}$ added to the time series.
            The data used for the estimation is shown in Fig.~\ref{fig1}(a).
            The Floquet exponents are estimated with $L = 1000$ and $t_{k+1} - t_{k} = 0.25$ in Eq.~\eqref{eq:lyapunov}.
            The natural frequency is estimated as $\omega \simeq 0.9997$ and the non-zero Floquet exponent is estimated as $\lambda \simeq -2.0457$, while their theoretical values are $\omega=1$ and $\lambda=-2$, respectively.
            The estimated value of the Floquet exponent is plotted with respect to the length $L$ in Fig.~\ref{fig1}(b).
            The maximum degree of the polynomial is set to $p = 18$.

            We estimated the phase function ${\Theta}$ by the proposed method (Sec.~\ref{sec3B}) from the time series data (Fig.~\ref{fig1}(a)). 
            The estimated phase function (Fig.~\ref{fig1}(c)) is compared with the true one (Fig.~\ref{fig1}(d)).
            The phase function is accurately estimated by the proposed method except for the central and peripheral regions far from the limit cycle. 
            The discrepancy is due to the lack of data in those regions (trajectories stay near the limit-cycle attractor most of the time) and also due to the limited representation power of the polynomials.

			We evaluated the estimation accuracy of the PSFs and PRFs.
			Figure~\ref{fig2} compares the PSFs $Z_1$ and $Z_2$ obtained from the estimated phase function using Eq.~(\ref{eq:estZ}) with the true function obtained by solving the adjoint equation (\ref{eq:adjZ}). 
			Figure~\ref{fig3} compares the nPRFs $G_{\varsigma,j}$ for finite-intensity impulses obtained from the estimated phase function using Eq.~(\ref{eq:estG}) with the true functions obtained from the analytical solution.
			The estimated functions are well consistent with the true ones and the estimation error is small; the accuracy of the estimation is $R_{Z}^{2} \geq 0.9859$ for the PSFs and $R_{G}^{2} \geq 0.9903$ for the nPRFs, respectively (Table~\ref{tab:SL_phase_error}). 
			These results indicate that the proposed method accurately estimates the phase function in the vicinity of the limit-cycle and also a bit farther out from the limit cycle where moderate nonlinearity comes in.

            Next, we estimated the amplitude function $R$ by the proposed method (Sec.~\ref{sec3C}) from the same time series data (Fig.~\ref{fig1}(a)). 
            The hyper parameter $\gamma$ is determined to be $\gamma = 1.0 \times 10^6$ from the L-curve. 
            The estimated amplitude function (Fig.~\ref{fig1}(e)) is compared with the true amplitude function (Fig.~\ref{fig1}(f)).
            Our method can accurately estimate the amplitude function near the limit cycle, but the estimate deviates in the central and peripheral regions. 
            In particular, the estimate cannot reproduce the divergence of the amplitude function near the {fixed point}. 
            This discrepancy is due to the lack of data and the singularity near the fixed point of the system.

\begin{figure}[tb]
                \centering
                \includegraphics[width=\hsize]{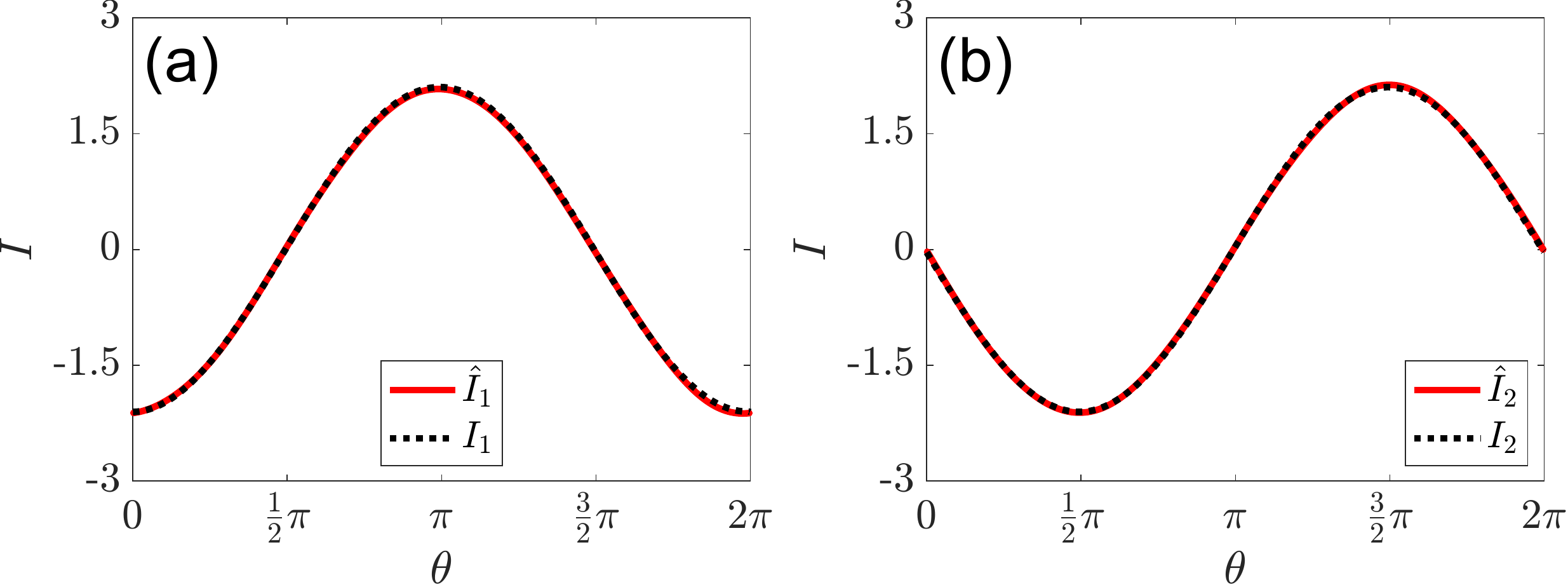}
                \caption{
                    Amplitude sensitivity functions (ASFs) of the SL oscillator.
                    Each graph shows the estimated ASF (red solid curve) and the true ASF (black dotted curve).
                    (a)~$x_1$ component.
                    (b)~$x_2$ component.
                    Estimated ASFs are obtained by numerical differentiation with $\epsilon=10^{-5}$. True ASFs are calculated by solving the adjoint equation.
                }
                \label{fig4}
            \end{figure}

\begin{figure}[tb]
                \centering
                \includegraphics[width=\hsize]{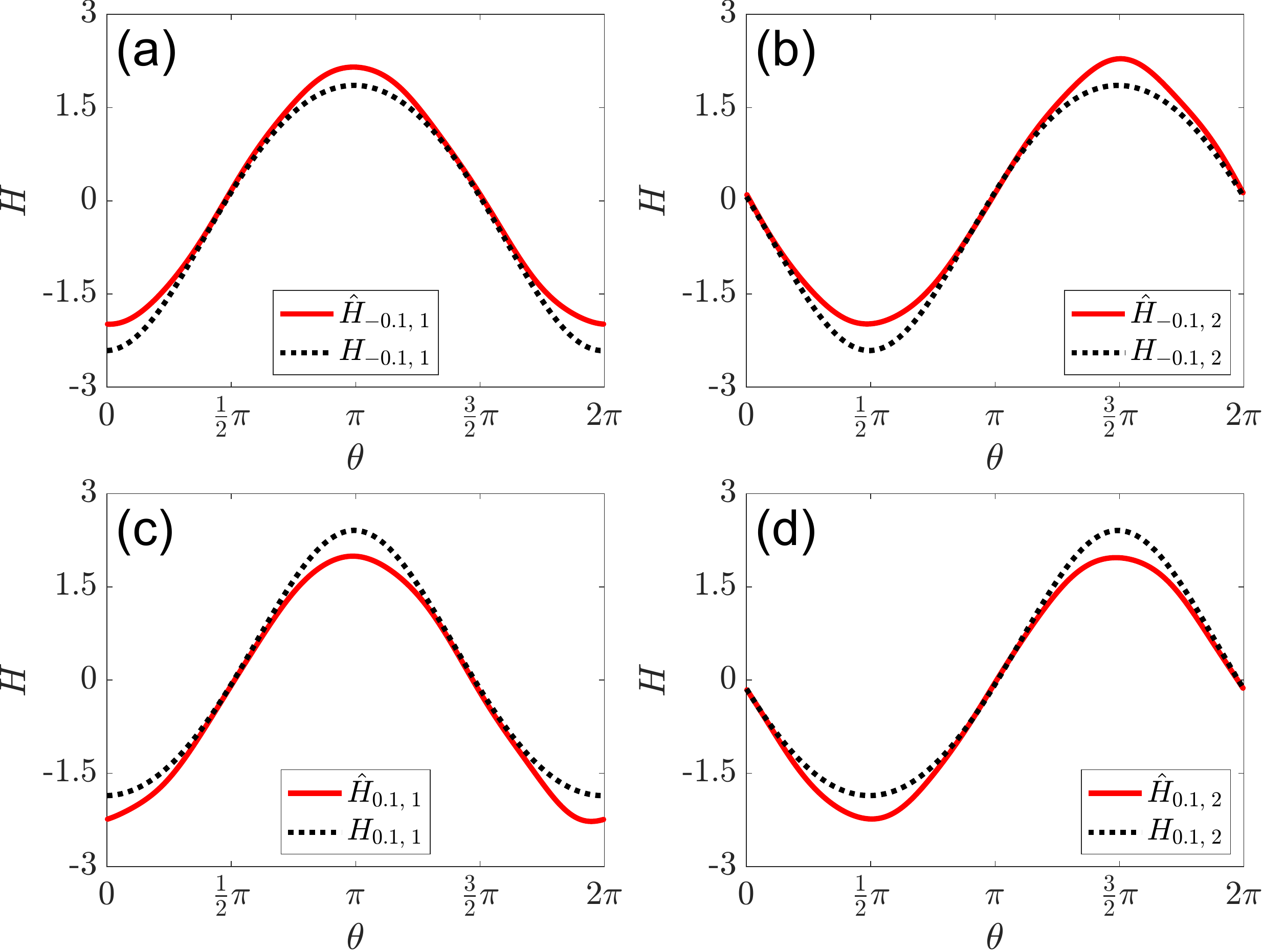}
                \caption{
                    Normalized amplitude response functions (nARFs) of the SL oscillator.
                    Each graph shows the estimated nARF (red solid curve) and the true nARF (black dotted curve).
                    (a,c):~Impulses applied in the $x_1$ direction.
                    (a)~$\varsigma = -0.1$,
                    (c)~$\varsigma = +0.1$.
                    (b,d):~Impulses applied in the $x_2$ direction.
                    (b)~$\varsigma = -0.1$,
                    (d)~$\varsigma = +0.1$.
                }
                \label{fig5}
            \end{figure}

			Figure~\ref{fig4} compares the ASFs $I_{1}$ and $I_{2}$ obtained from the estimated amplitude function using Eq.~(\ref{eq:estI}) with the true function obtained by solving the adjoint equation~(\ref{eq:adjI}).
			Figure~\ref{fig5} compares the nARFs $H_{\varsigma,j}(\theta)$ obtained from the estimated amplitude function using Eq.~(\ref{eq:estH}) with the true results.  
            Despite the discrepancy in the amplitude function far from the limit cycle, the estimated results agree reasonably well with the true results for the impulse intensities used here.
            The accuracy of estimation is $R_{I}^{2} \geq 0.9998$ for the ASFs and $R_{H}^{2} \geq 0.9725$ for the nARFs (Table~\ref{tab:SL_amp_error}).
            These results confirm that the amplitude function is estimated appropriately around the limit cycle including moderately nonlinear regimes.

%
%
            \begin{table}[h]
                \centering
                \caption{Coefficients of determination of the nPRFs and PSFs of the SL oscillator.}
                \label{tab:SL_phase_error}
            \begin{tabular}{cccccc}\hline \hline
                $Z_{1}$ & $Z_{2}$ & $G_{-0.2,1}$ & $G_{-0.2,2}$ & $G_{0.2,1}$ & $G_{0.2,2}$ \\
                $0.9869$ & $0.9859$ & $0.9912$ & $0.9903$ & $0.9927$ & $0.9929$ \\ \hline \hline
            \end{tabular}
			%
                \centering
                \caption{Coefficients of determination of the nARFs and ASFs of the SL oscillator.}
                \label{tab:SL_amp_error}
           \begin{tabular}{cccccc}\hline \hline
                $I_{1}$ & $I_{2}$ & $H_{-0.1,1}$ & $H_{-0.1,2}$ & $H_{0.1,1}$ & $H_{0.1,2}$ \\
                $0.9998$ & $0.9999$ & $0.9779$ & $0.9727$ & $0.9725$ & $0.9742$ \\ \hline \hline
            \end{tabular}
            \end{table}


        \subsection{van der Pol oscillator}

            Next, we verify the validity of the proposed method using the van der Pol (vdP) oscillator.
            The vdP oscillator~\cite{van1926lxxxviii,van1927vii,van1927frequency} is described as follows:
            \begin{align}
				\label{eq:vdp}
                \ddt
                \begin{bmatrix}
                    x_{1} \\ x_{2}
                \end{bmatrix}
                = 
                \begin{bmatrix}
                    x_{2} \\
                    \nu \left( 1 - x_{1}^{2} \right)x_{2} - x_{1}
                \end{bmatrix}
                ,
            \end{align}
            where $x_1, x_2$ are the variables and the parameter is chosen as $\nu = 1$.
            We set the sampling interval as $\Dt = 0.005$, the number of initial points as $n = 1200$, and the number of data points as $M = 1000$.
            We estimated the phase function for the vdP oscillator under the observation noise, given by
            Gaussian noise with mean $0$ and standard deviation $5 \times 10^{-3}$ to the time series as the observation noise.
            The data used for the estimation is shown in Fig.~\ref{fig6}(a).
            The Floquet exponents are estimated with $L = 1000$ and $t_{k+1} - t_{k} = 0.25$ in Eq.~\eqref{eq:lyapunov}.
            The natural frequency is estimated as $\omega \simeq 0.9434$ and the non-zero Floquet exponent is estimated as $\lambda \simeq -1.0885$. 
            The latter value agree well with the theoretical value $\lambda \simeq -1.0581$ that is evaluated from the monodromy matrix of the original model~\cite{Takata2021}.
            The estimated value of the Floquet exponent with respect to the length $L$ is shown in Fig.~\ref{fig6}(b).
            The maximum degree of the polynomial is set to $p = 18$.

            We estimated the phase function ${\Theta}$ by the proposed method (Sec.~\ref{sec3B}) from the time series data (Fig.~\ref{fig6}(a)). 
            The estimated result (Fig.~\ref{fig6}(c)) is compared with the true phase function (Fig.~\ref{fig6}(d)) obtained directly from the mathematical model, Eq.~(\ref{eq:vdp}).
			As in the case with the SL oscillator, the estimated function is close to the true one except for the central and peripheral regions far from the limit cycle.

			We examined the estimation accuracy of the PSFs and PRFs.
			Figure~\ref{fig7} compares the estimated PSFs $Z_1,$ and $Z_2$ with the true ones.
            The estimated PSFs agree well with the true PSFs, implying the accurate estimation of the phase function in the vicinity of the limit cycle.
            Figure~\ref{fig8} compares the estimated nPRFs $G_{\varsigma,j}(\theta)$ for finite-intensity impulses with the true results. 
            Again, the estimated nPRFs agree well with the true results. 
            Overall, the proposed method can estimate the PSFs and nPRFs accurately (Table~\ref{tab:vdP_phase_error}): $R_{Z}^{2} \geq 0.9896$ and $R_{G}^{2} \geq 0.9892$ for PSF and nPRFs, respectively. 
            These results indicate that the proposed method accurately estimates the phase function in the moderately nonlinear regime around the limit cycle.
            
			Next, we estimated the amplitude function $R$ by the proposed method (Sec.~\ref{sec3C}) from the same data (Fig.~\ref{fig6}(a)). 
            The hyper parameter $\gamma$ is determined to be $\gamma = 1.0 \times 10^3$ from the L-curve. 
            The estimation result (Fig.~\ref{fig6}(e)) is compared with the true amplitude function (Fig.~\ref{fig6}(f)) obtained directly from the mathematical model, Eq.~({\ref{eq:vdp}).
			As in the SL case, the estimate agrees with the true function near the limit cycle, but a discrepancy arises in the central and peripheral regions far from the limit cycle. 

			We examined the estimation accuracy of the ASFs and ARFs. 
            Figure~\ref{fig9} compares the estimated ASFs with the true ASFs, showing a reasonable agreement. 
            Figure~\ref{fig10} compares the estimated nARFs with the true nARFs for finite-intensity impulses. 
            The estimates also agree well with the true nARFs. 
            Again, the proposed method can estimate the ASF and nARFs accurately (Table~\ref{tab:vdP_amp_error}): $R_{I}^{2} \geq 0.9792$ and $R_{H}^{2} \geq 0.9694$ for ASF and nARFs, respectively. 
            These results suggest that the proposed method estimates the amplitude function around the limit cycle accurately.


            \begin{figure}[tb]
                \centering
                \includegraphics[width=\hsize]{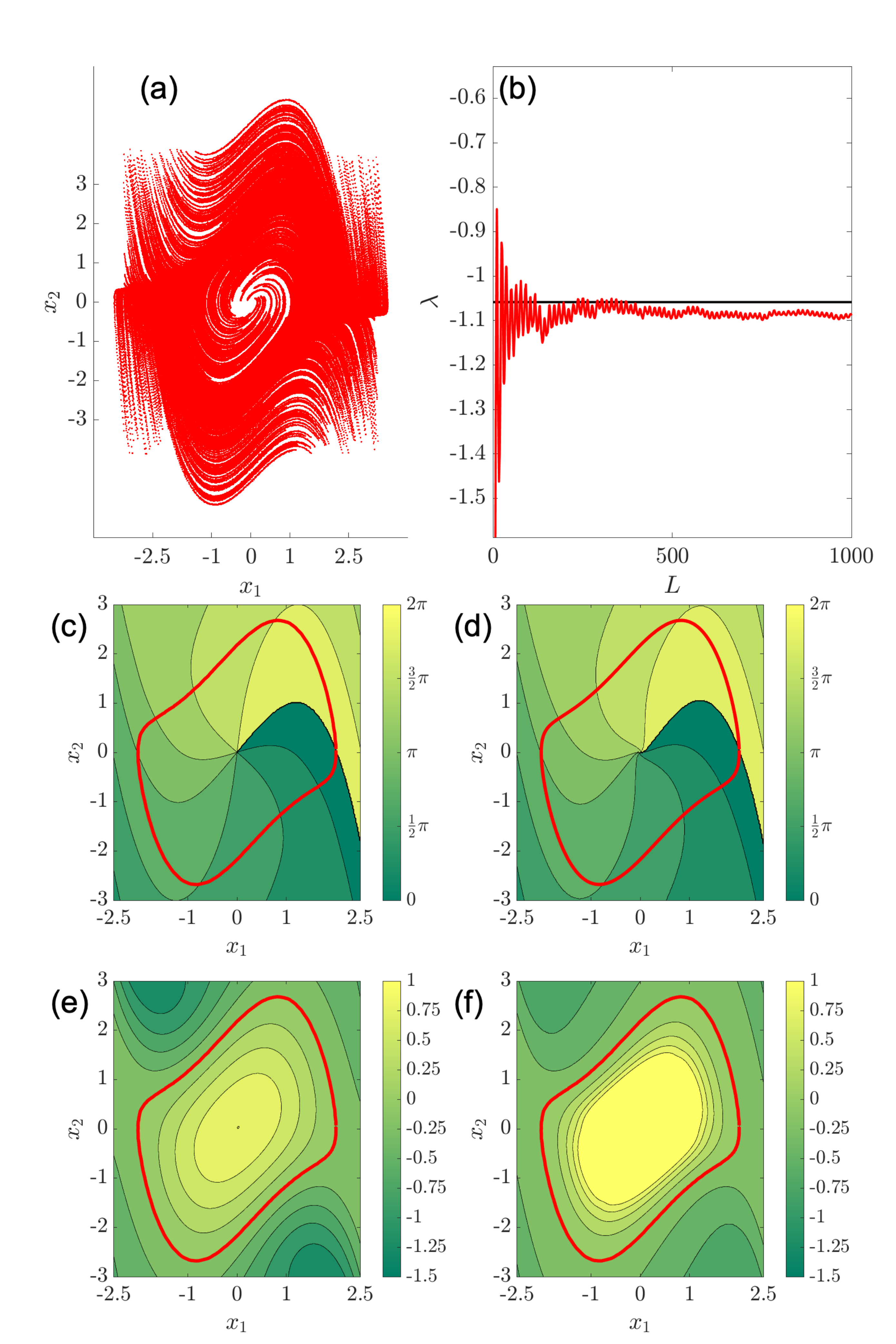}
                \caption{
				    Phase and amplitude functions of the vdP oscillator. 
                    (a)~Time series data of the vdP oscillator used for the estimation. 
                    (b)~Estimated non-zero Floquet exponent.
                    (c)~Estimated phase function. 
                    (d)~True phase function. 
                    (e)~Estimated amplitude function. 
                    (f)~True amplitude function. 
                    In (c)-(f), the red curve shows the limit cycle.
                }
                \label{fig6}
            \end{figure}
            \begin{figure}[tb]
                \centering
                \includegraphics[width=\hsize]{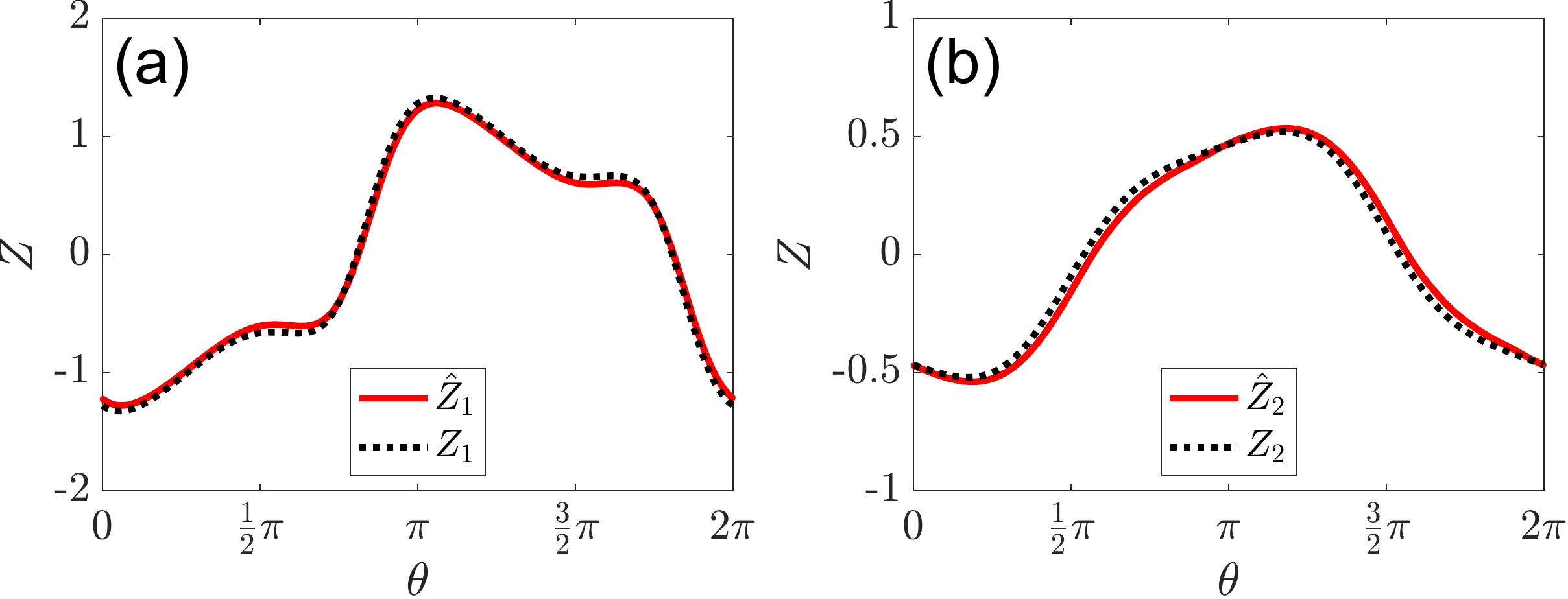}
                \caption{
                    Phase sensitivity functions (PSFs) of the vdP oscillator.
                    Each graph shows the estimated PSF (red solid curve) and the true PSF (black dotted curve).
                    (a): $x_1$ component.
                    (b): $x_2$ component.
                    Estimated PSFs are obtained by numerical differentiation with $\epsilon=10^{-5}$. True PSFs are calculated by solving the adjoint equation.
                }
                \label{fig7}
            \end{figure}
            \begin{figure}[tb]
                \centering
                \includegraphics[width=\hsize]{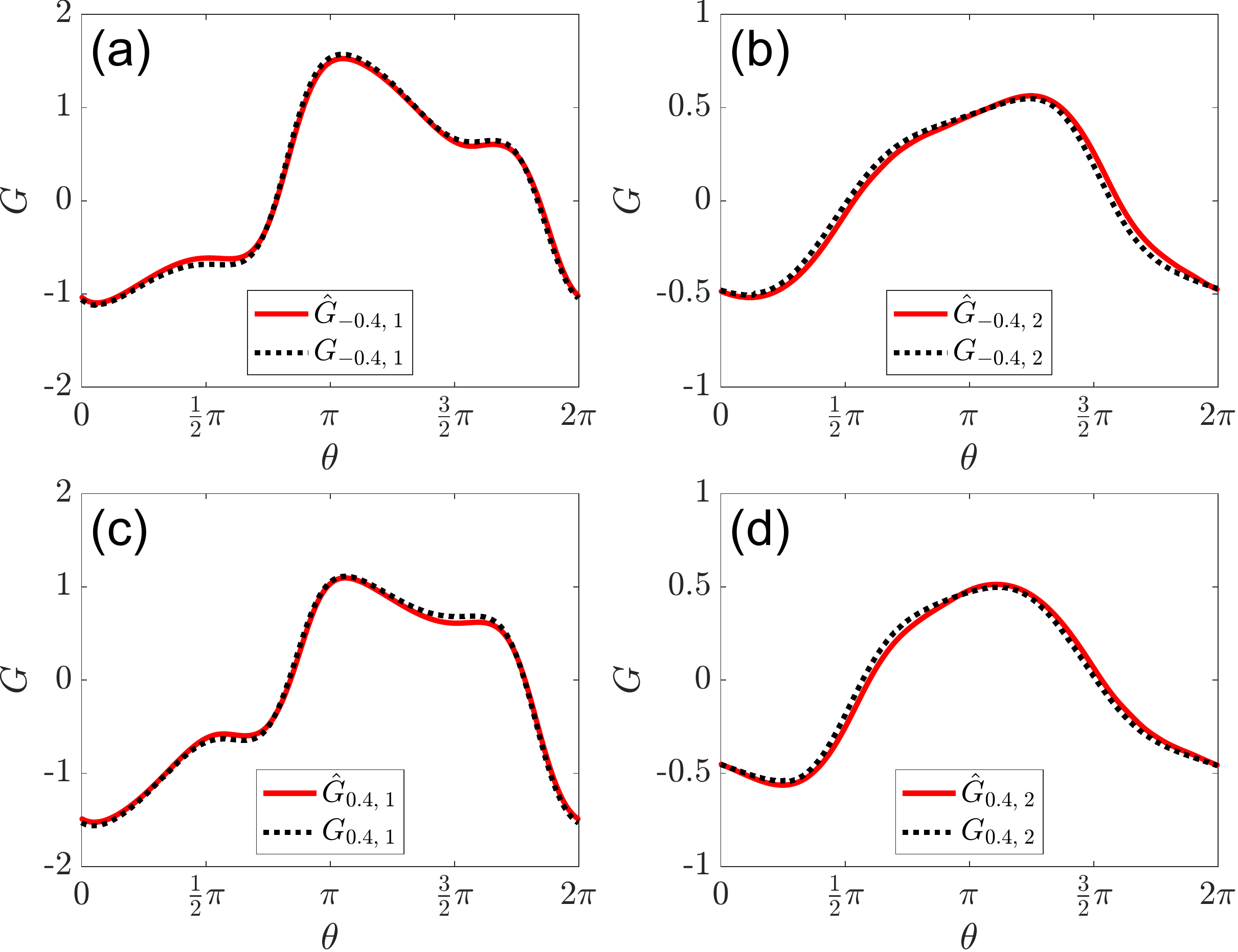}
                \caption{
					Normalized phase response functions (nPRFs) of the vdP oscillator. 
                    Each graph shows the estimated nPRF (red solid curve) and the true nPRF (black dotted curve).
                    (a,c):~Impulses applied in the $x_1$ direction.
                    (a)~$\varsigma = -0.4$,
                    (c)~$\varsigma = +0.4$.
                    (b,d):~Impulses applied in the $x_2$ direction.
                    (b)~$\varsigma = -0.4$,
                    (d)~$\varsigma = -0.4$.
                }
                \label{fig8}
            \end{figure}

            \begin{figure}[tb]
                \centering
                \includegraphics[width=\hsize]{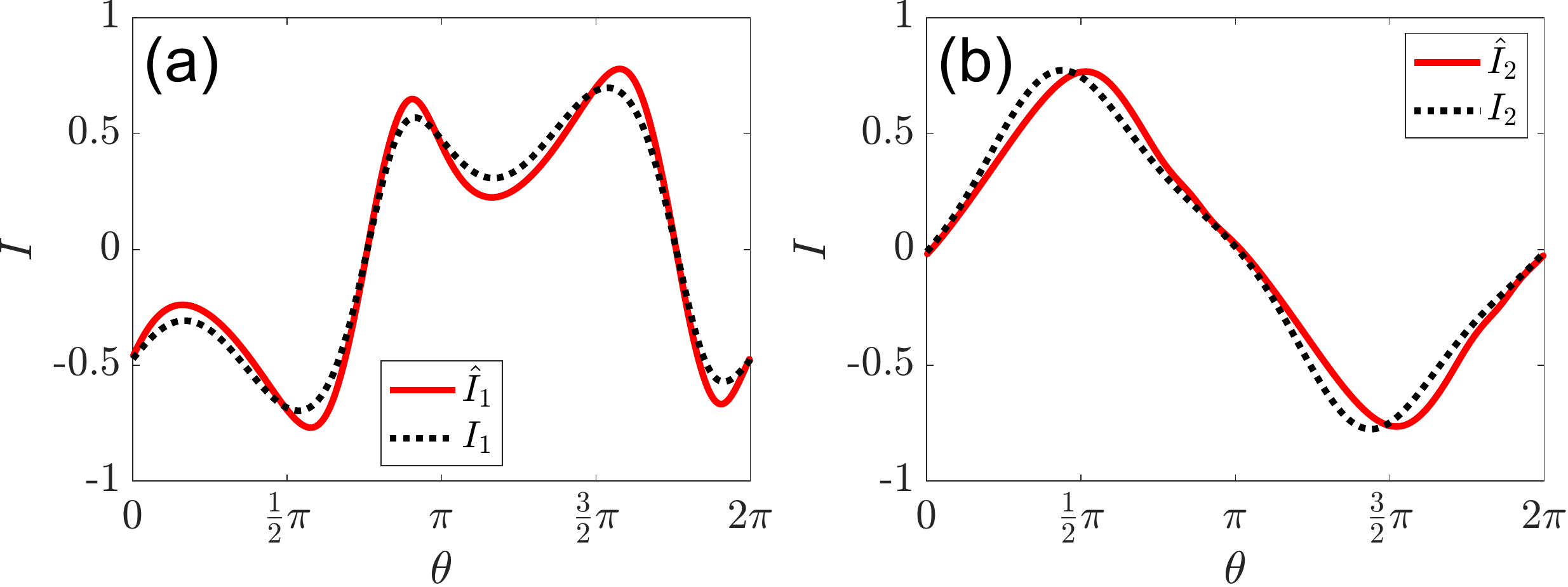}
                \caption{
					Amplitude sensitivity functions (ASFs) of the vdP oscillator. 
                    Each graph shows the estimated ASF (red solid curve) and the true ASF (black dotted curve).
                    (a): $x_1$ component.
                    (b): $x_2$ component.
                    Estimated ASFs are obtained by numerical differentiation with $\epsilon=10^{-5}$. True ASFs are calculated by solving the adjoint equation.
                }
                \label{fig9}
            \end{figure}
            \begin{figure}[tb]
                \centering
                \includegraphics[width=\hsize]{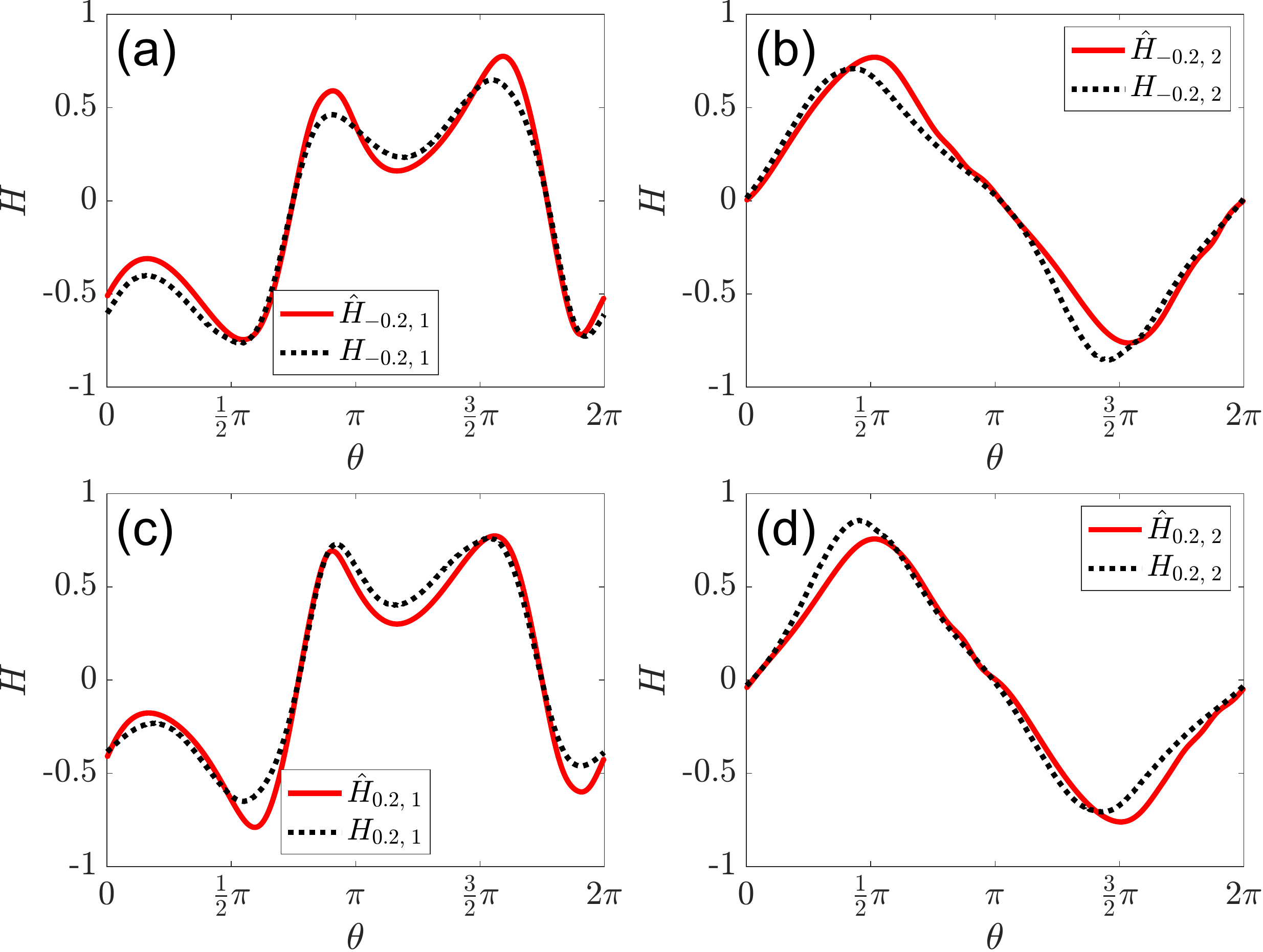}
                \caption{
                    Normalized amplitude response functions (nARFs) of the vdP oscillator.
                    Each graph shows the estimated nARF (red solid curve) and the true nARF (black dotted curve).
                    (a,c):~Impulses applied in the $x_1$ direction.
                    (a)~$\varsigma = -0.2$,
                    (c)~$\varsigma = +0.2$.
                    (b,d):~Impulses applied in the $x_2$ direction.
                    (b)~$\varsigma = -0.2$,
                    (d)~$\varsigma = +0.2$.
                }
                \label{fig10}
            \end{figure}


            \begin{table}[h]
                \centering
                \caption{Coefficients of determination of the nPRFs and PSFs of the vdP oscillator.}
                \label{tab:vdP_phase_error}
            \begin{tabular}{cccccc}\hline \hline
                $Z_{1}$ & $Z_{2}$ & $G_{-0.4,1}$ & $G_{-0.4,2}$ & $G_{0.4,1}$ & $G_{0.4,2}$ \\
                $0.9971$ & $0.9892$ & $0.9980$ & $0.9896$ & $0.9979$ & $0.9914$ \\ \hline \hline
            \end{tabular}
            %
                \centering
                \caption{Coefficients of determination of the nARFs and ASFs of the vdP oscillator.}
                \label{tab:vdP_amp_error}
            \begin{tabular}{cccccc}\hline \hline
                $I_{1}$ & $I_{2}$ & $H_{-0.2,1}$ & $H_{-0.2,2}$ & $H_{0.2,1}$ & $H_{0.2,2}$ \\
                $0.9795$ & $0.9792$ & $0.9753$ & $0.9694$ & $0.9707$ & $0.9736$ \\ \hline \hline
            \end{tabular}
            \end{table}

    \section{Data-driven optimal entrainment with amplitude suppression}
    \label{sec5}

		In this section, we apply the proposed method to the optimal entrainment of the oscillator with amplitude suppression using the PSF and ASF developed in Ref.~\cite{Takata2021}.
        We consider a limit-cycle oscillator driven by an external periodic input. 
        The natural frequency of the oscillator is $\omega$ and the frequency of the periodic input is $\Omega$, where $\Omega$ is close to $\omega$.
        When the input is sufficiently weak, the reduced and averaged phase equation for the phase difference $\phi$ between the oscillator and the external periodic input is given by
        \begin{align}
            \ddt \phi = \Delta + \Gamma(\phi), \quad \Gamma(\phi) = \left[ \bZ(\phi + \Omega t) \cdot \bq(\Omega t) \right]_{t},
        \end{align}
        where $\Delta = \omega - \Omega$ is the frequency mismatch between the periodic input and the oscillator, $\Gamma$ is the phase coupling function, $\bq$ is the periodic input, and $\left[ f(t) \right]_{t} =(\Omega/2\pi) \int_0^{2\pi/\Omega} f(s) ds$ denotes the average of $f$ over one period of the external input.
        The optimal periodic input $\bq$ can be obtained by solving the following optimization problem as discussed in Ref.~\cite{Zlotnik2013}:
        \begin{align}
            \label{eq:opt_input}
            \begin{aligned}
                &&\bq ={} & \argmax_{\bq}\; - \Gamma'(\phi^{*}) \\
                &&&\mathrm{s.t.}\quad \Delta + \Gamma(\phi^{*}) = 0,\; \left[ \| \bq \|^{2} \right]_{t} = Q,
            \end{aligned}
        \end{align}
        where $\Gamma'$ is the derivative of the phase coupling function at the stable phase locking point $\phi^{*}$;
        the first constraint is for the system to have a phase-locked solution at $\phi^*$, and the second constraint is for the power of the periodic input to be $Q > 0$.
        However, when the periodic input is not sufficiently weak, the trajectory deviates from the limit cycle and the above method based on the phase reduction may not work appropriately.

        In the amplitude-penalty method~\cite{Takata2021}, the amplitude equation is also used to suppress the deviations of the trajectory from the limit cycle.
        It is formulated by adding a penalty term characterizing the effect of the input on the amplitude variable of the oscillator to the optimization problem~\eqref{eq:opt_input} as follows:
        \begin{align}
            \label{eq:penalty_opt}
            \begin{aligned}
                &&\bq ={} & \argmax_{\bq}\; - \Gamma'(\phi^{*}) - k \left[ \left( \bI(\phi^{*} + \Omega t) \cdot \bq( \Omega t) \right)^{2} \right]_{t} \\
                &&&\mathrm{s.t.}\; \Delta + \Gamma(\phi^{*}) = 0,\; \left[ \| \bq \|^{2} \right]_{t} = Q,
            \end{aligned}
        \end{align}
        where the second term in the objective function represents the penalty for the amplitude deviation and $k > 0$ is the weight of the penalty.
        By solving the modified optimization problem~\eqref{eq:penalty_opt}, the optimal input $\bq$ that simultaneously improves the stability of the phase-locked solution $\phi^*$ and suppresses the deviation of the amplitude from the limit cycle can be obtained.

        The results of the amplitude penalty method for the vdP oscillator using the PSF and ASF estimated under the observation noise are shown in Fig.~\ref{fig11}.
        The parameters are set to $k = 2 \times 10^{4}, Q = 0.05$, $\phi^* = 0$, and $\Delta = 0$.

        Figure~\ref{fig11}(a) shows the evolution of the phase differences with and without the amplitude penalty, where results using the PSF and ASF calculated by the adjoint method (theoretical) and those estimated from the observed data (data-driven) are compared.
        The amplitude penalty method leads to the correct phase-locking point $\phi^*=0$, where the results using both PSFs and ASFs are almost indistinguishable, while the conventional method without amplitude suppression leads to incorrect phase-locking point different from $\phi^* = 0$ due to amplitude deviations.

        Figure~\ref{fig11}(b) shows the trajectories driven by the periodic inputs with and without amplitude penalty, where the estimated PSF and ASF are used.
        It can be seen that the trajectory driven by the periodic input with amplitude penalty is almost the same as the limit cycle without perturbation, while the trajectory without amplitude penalty deviates from the limit cycle. This result suggests that the proposed method is a promising approach to realize the optimal entrainment of the oscillator in a data-driven manner without the knowledge of the mathematical model.

        \begin{figure}[tb]
            \centering
            \includegraphics[width=\hsize]{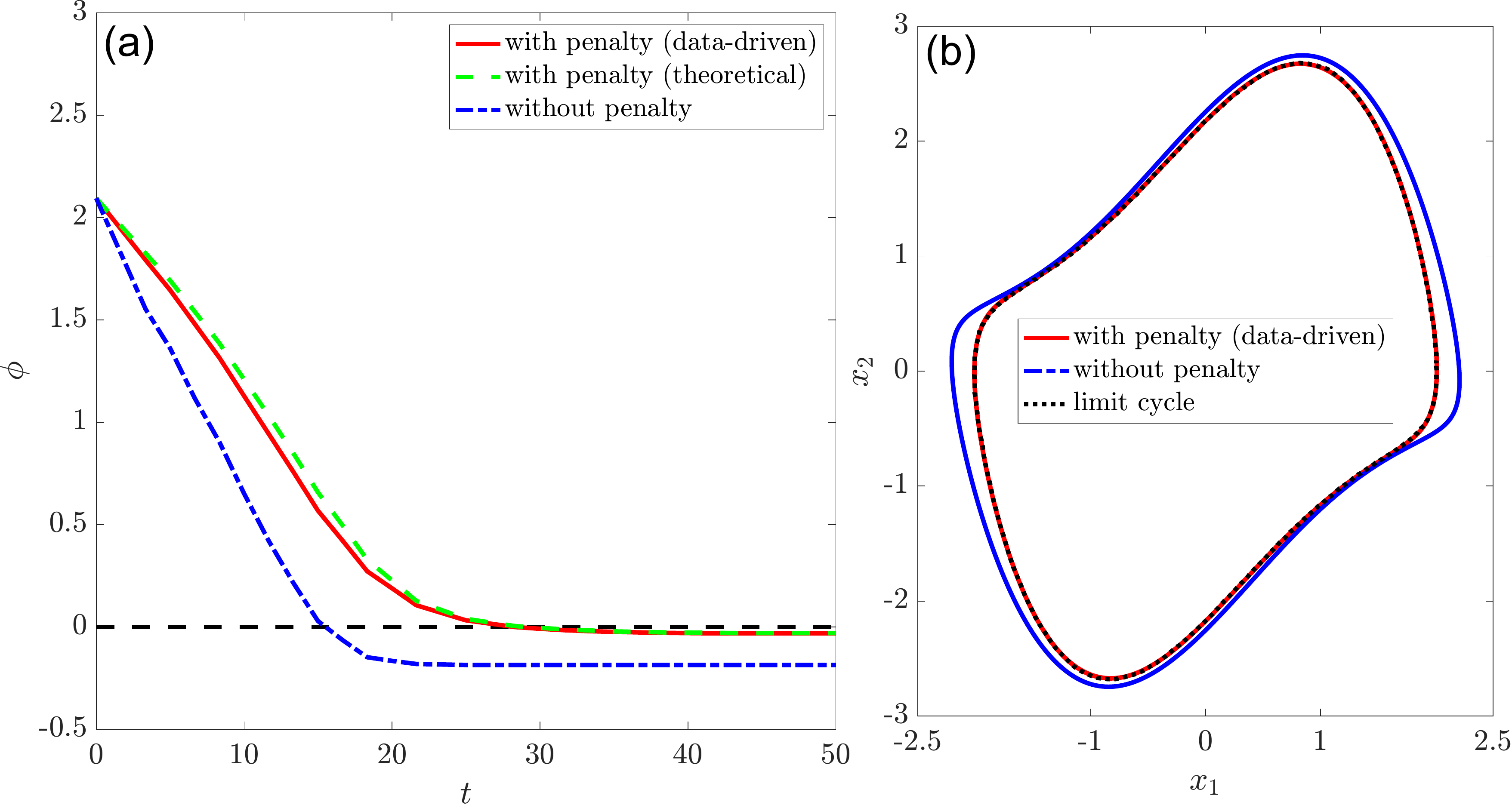}
            \caption{
                Data-driven optimal entrainment of the vdP oscillator.
                (a)~Evolution of the phase differences for the case with periodic input obtained from the estimated PSF and ASF (red solid), for the case with input obtained from true PSF and ASF (green dotted), and for the case  without amplitude suppression (blue dashed).
                (b)~Trajectories of the oscillator state; the case with periodic input obtained from the estimated PSF and ASF (red solid) and the case without the amplitude penalty (blue dashed) are compared. Black dotted curve shows the limit cycle without periodic input.
                }
            \label{fig11}
        \end{figure}

    \section{Concluding remarks} 
        \label{sec6}

        In this study, we proposed a method for estimating the phase and amplitude functions of limit-cycle oscillators only from observed time series. 
        Our method is based on the polynomial regression, which is simple and computationally efficient. 
        The resulting optimization problems are convex and can be easily solved. The numerical results based on two limit-cycle oscillators (SL and vdP oscillator) showed that our method can estimate the phase and amplitude function accurately around the limit cycle under the observation noise. 
        Furthermore, we demonstrated that the proposed method enables us to achieve the optimal entrainment with amplitude suppression in a data-driven manner, and the performance was comparable to the model-based case.

	    The proposed method cannot accurately estimate the phase and amplitude functions in the phase-plane regions away from the limit cycle (Fig.~\ref{fig1} and \ref{fig6}). 
        In particular, it cannot estimate the singular behavior of the amplitude function near the {fixed point} of the system.
        However, our method accurately estimates the PSF and nPRFs (Figs. 2, 3, 7, and 8) and ASF and nARFs (Figs. 4, 5, 9, and 10). 
        In the phase-amplitude description, the PSF, nPRFs, ASF, and nARFs are essential functions for the phase-amplitude description of limit-cycle oscillators.  
        Thus, the discrepancy of the phase and amplitude functions away from the limit cycle is not a serious problem for applications. 
        Indeed, we demonstrated that the proposed method is useful for data-driven optimal entrainment of the limit-cycle oscillators (Fig. 11). 
        The proposed method is potentially helpful also for data-driven analysis and control of limit-cycle oscillators subjected to pulsatile forcing or coupling~\cite{Ermentrout2010,Nakao2005,arai2008phase}.

        As mentioned in the introduction, a major alternative to our proposed method is the EDMD~\cite{Williams2015}, which estimates Koopman eigenvalues and eigenfunctions from the observed data.
		The asymptotic phase function $\Theta$ and amplitude function $R$ are related to the Koopman eigenfunctions with Koopman eigenvalues $i\omega$ and $\lambda$ respectively, namely, they satisfy $(d/dt) e^{i \Theta} = i\omega e^{i \Theta}$ and $(d/dt) R = \lambda R$~\cite{mauroy2016global,mauroy2018global,Mauroy2020,mauroy2013isostables,Shirasaka2017}.
        The primary difference of the present study from those based on EDMD is that the natural frequency $\omega$ and the Floquet exponent $\lambda$ are given as external parameters (measured separately) and then the phase and amplitude functions are estimated, while in the EDMD, the frequency and Floquet exponents are estimated from the time series together with the eigenfunctions. 
        Therefore, our method can be more stable, because $\omega$ and $\lambda$ are measured separately by using the classical, well-established method for the estimation of Lyapunov exponents~\cite{lyapunov}.
        We thus think that our proposed method is complementary to the EDMD methods in the analysis of real-world oscillatory signals.

        In this study, we applied the proposed method to limit-cycle oscillators with known mathematical models in order to evaluate its accuracy.
        Our future work is to apply the method to real-world observed data such as ECGs~\cite{heart}.
        To this end, we plan to further develop a method of preprocessing for the time series to mitigate the effect of stronger noise and non-stationarity.


	\acknowledgments
    H. N. thanks financial support from JSPS KAKENHI (Nos. JP17H03279, JP18H03287, and JPJSBP120202201) and JST CREST (No. JP-MJCR1913). R. K. thanks  JSPS KAKENHI (Nos. 18K11560, 19H01133, 21H03559, 21H04571), JST PRESTO (No. JPMJPR1925), and AMED (No. JP21wm0525004).
	

%
    
\end{document}